\documentstyle[titlepage,fleqn,12pt]{article}
\begin{document}
\begin{titlepage}
\title{Singularity spectrum of self-organized criticality}
\author{E. Canessa \\
\\
ICTP-International Centre for Theoretical Physics, Trieste,
Italy}
\date{}
{\baselineskip=25pt

\begin{abstract}

I introduce a simple continuous probability theory based on
the Ginzburg-Landau equation that provides for the first time
a common analytical basis to relate and describe the main features
of two seemingly different phenomena of condensed-matter physics,
namely self-organized criticality and multifractality.
Numerical support is given by a comparison with reported simulation
data.  Within the theory the origin of self-organized critical
phenomena is analysed in terms of a nonlinear singularity spectrum
different from the typical convex shape due to multifractal measures.

\vskip 2cm

\end{abstract}
    }
\maketitle
\end{titlepage}
\baselineskip=25pt
\parskip=0pt

The concept of self-organized criticality (SOC) \cite{Bak87}
has attracted great interest recently, both analytically \cite{Lee91,Gri90}
and experimentally \cite{Dio91,Cot91}.
The idea behind SOC is that a certain class of dynamical many-body
systems drive themselves into a statistically stationary critical state,
with no intrinsic length or time scale, where they exhibit fractal
behaviour and generate $1/f$ noise.
Besides SOC, the generalization of fractal growth to
self-similar multifractals has also attracted considerable attention from
the physics community over the past years (see, {\em e.g.},
\cite{Am86,Lee88}).  Theoretical models to describe MF have been
concerned with mean-field arguments \cite{Mut83} and standard
renormalization group methods \cite{Gou83}.  These approaches, however,
do not allow for the direct analysis of the (parabolic-shaped) singularity
spectrum typical for a multifractal structure \cite{Am86,Lee88,Chh89}.

Motivated by the recent suggestion that SOC supports the appearence
of fractal structures \cite{Bak87}, it is natural to ask then if
there is a common principle underlying the
seemingly unrelated phenomena of SOC and MF.  So far as I know a fixed
scale transformation method \cite{Piwr91}, developed for fractal growth,
has been used to investigate analytically the nature of 2D clusters in SOC.
Henceforth, it is also tempting to search for a unifiying scenario that
underpins a plausible link between MF and SOC.  In fact this is the motivation
for this work in which I only take a step in that direction.

In this rapid communication I propose a simple continuous probability
theory based on the Ginzburg-Landau (GL) equation \cite{Gin50} that
combines together the concepts of SOC and MF for the first time.  In
this goal I explore a new analytical basis which, on the one hand, allows
to understand the genesis of SOC from the point of view of a nonlinear
singularity spectrum and, on the other hand, reveals further insight
into the physics governing this crossover.  In short, I make here an
attempt to unravel the basic phenomena at the origin of power laws
and multifractal dimensions.

A crucial feature of the present formalism is to consider that
{\em all random variables} in a one-dimensional (1D) space,
${\cal R}^{1}$, are functions of the coordinate variable
$\chi$ which I map into an equivalent independent
variable $\zeta$ -say, $energy/unit force$-
characterizing a random system.  Then, all probabilities
may became expressible in terms of the uniform
{\em probability distribution function} \cite{Fel66}
\begin{equation}\label{eq:can2}
{\cal G}(\zeta_{2} )-{\cal G}(\zeta_{1}) ={\bf P}
     \{ \zeta_{1}< {\bf \zeta} \leq \zeta_{2} \} \approx
     \int_{\zeta_{1}}^{\zeta_{2}}  \phi (\zeta )
       \; d\zeta    \;\;\; ,
\end{equation}
where $\{ \}$ indicates the function interval and $\phi$ is
a uniform {\em probability density} on the line (or ${\cal R}^{1}$)
which needs to be specified.

Assume, within this continuous probability model, that the order
parameter $\phi$ satisfies
\begin{equation}\label{eq:can2a}
\phi (\zeta ) \equiv \frac{\phi_{0}}{2}\{ 1+\mu H(\zeta ) \}  \;\;\; ,
\end{equation}
such that $\phi (\zeta \rightarrow + \infty )/\phi_{0}\rightarrow 0$
and $\phi (\zeta \rightarrow -\infty )/\phi_{0}\rightarrow 1$,
({\em i.e.}, $\mu H(+\infty )=-1$ and $\mu H(+\infty )=1$),
and $\mu$ is a coefficient as discussed below ({\em c.f.},
Eq.(\ref{eq:Madd2})).

I postulate $H(\zeta )$ in Eq.(\ref{eq:can2a}) to be given by the real
solutions of the following static, dimensionless GL-like equation \cite{Gin50}:
$\partial^{2} H(\zeta^{*})/\partial (\zeta^{*})^{2} +
 p H(\zeta^{*})-q H^{3}(\zeta^{*}) =0, \;\;\; \zeta^{*} \varepsilon  D$,
where $\zeta^{*}\equiv \zeta /\zeta_{o}$, $D$ is the $\zeta$-domain,
$[p,q]>0$ and $\zeta_{o}$ is a positive coefficient of dim[$length$].
It is not unreasonable to consider the (1D) domain $D$ to be
infinite.  Then, the GL posseses the well-known stable, kink solution
\begin{equation}
H(\zeta^{*})=  \pm \sqrt{\frac{p}{q}} \;
    tanh (\zeta^{*}\sqrt{\frac{p}{2}}) \;\;\; .
\end{equation}
Using this result together with Eqs.(\ref{eq:can2})
and (\ref{eq:can2a}), I will establish a simple relation
for the probability distribution ${\bf P}$.

The integral of Eq.(\ref{eq:can2}) over the limits:
$\zeta_{2} \equiv \lambda_{1} \zeta_{o} \geq \zeta + \lambda_{2}
\zeta_{o}\equiv \zeta_{1}$ is
\begin{equation}\label{eq:cca1}
 {\cal G}(\lambda_{1} \zeta_{o}) - {\cal G}(\zeta + \lambda_{2} \zeta_{o})
    =  \frac{\phi_{0}}{2}\int_{\zeta + \lambda_{2}
            \zeta_{o}}^{\lambda_{1} \zeta_{o}}
 \{ 1 \pm \mu \; tanh (\frac{\zeta '}{\zeta_{o}}) \} \; d\zeta '
      \equiv -\tau (\zeta )  \;\;\; ,
\end{equation}
where I introduced the new function $\tau (\zeta )$ and
set $p=q=2$ to reduce the number of free parameters.
These integration limits lead to the condition
\begin{equation}\label{eq:Ant1}
\lambda_{2} -\lambda_{1} +  \zeta^{*} \leq 0 \;\;\; .
\end{equation}
Note that the sign in Eq.(\ref{eq:cca1}) implies that the single functions
${\cal G}$ are here assumed to satisfy the condition
${\cal G}(\zeta + \lambda_{2} \zeta_{o}) > {\cal G}(\lambda_{1}\zeta_{o})$
for $\zeta \neq 0$, which throughout the theory are undefined, whereas the
free parameters $\lambda_{i}$ ($i$=1,2) will restrict the range of $\zeta^{*}$
via Eq.(\ref{eq:Ant1}).

Suppose $\lambda_{2}\neq \lambda_{1}$, then it is straightforward
to solve the above integral and get
\begin{eqnarray}\label{eq:w1}
\tau (\zeta^{*}) & \approx & (1+\frac{\zeta^{*}}{\lambda_{2} -\lambda_{1}})
    \{ \tau (0)\mp \mu\phi_{0}^{*} \ln cosh \lambda_{2}\}  \nonumber \\
        &   & \hspace{2.2cm} \pm
 \mu \phi_{0}^{*}  \{ \frac{\zeta^{*} }
      {\lambda_{2} -\lambda_{1}}\ln cosh \lambda_{1}
      + \ln cosh (\lambda_{2}+\zeta^{*})  \}   \;\;\; ,
\end{eqnarray}
in which $\phi_{0}^{*}\equiv \zeta_{o}\phi_{0}/2$ and
\begin{equation}\label{eq:Mat2}
\tau (0)  \equiv  {\cal G}(\lambda_{2} \zeta_{o})
      -{\cal G}(\lambda_{1} \zeta_{o})
     =  (\lambda_{2}-\lambda_{1}) \phi_{0}^{*} \{ 1\mp
     \mu \Gamma_{\lambda}  \}   \;\;\; ,
\end{equation}
such that
\begin{equation}\label{eq:Madd9}
\Gamma_{\lambda} \equiv \frac{ \ln cosh \lambda_{1} - \ln cosh \lambda_{2} }
    {\lambda_{2}-\lambda_{1}}      \;\;\; .
\end{equation}
When $\zeta^{*}=0$, MF will be described via Eq.(\ref{eq:Mat2}) assuming
${\cal G}(\lambda_{2} \zeta_{o})<{\cal G}(\lambda_{1} \zeta_{o})$, whereas
for SOC: ${\cal G}(\lambda_{2} \zeta_{o})>{\cal G}(\lambda_{1} \zeta_{o})$.

To analyse the possible multifractal features of {\bf P}
let me define the function $D_{\zeta^{*}}$ as
\begin{equation}\label{eq:w2}
\tau (\zeta^{*}) \equiv \{ \lambda_{2}-\lambda_{1}+\zeta^{*}\}  D_{\zeta^{*}}
    \;\;\; .
\end{equation}
{}From Eqs.(\ref{eq:w1}) and (\ref{eq:w2}) it follows that
\begin{eqnarray}\label{eq:Mat1}
D_{\zeta^{*}} & \equiv & \frac{1}{\lambda_{2}-\lambda_{1}}\{
   \tau (0)\mp \mu\phi_{0}^{*}\ln cosh \lambda_{2} \}  \nonumber \\
   & & \hspace{1.8cm} \pm \; \frac{\mu \phi_{0}^{*}}
       {\lambda_{2}-\lambda_{1}+\zeta^{*}}  \{ \frac{\zeta^{*}  }
       {\lambda_{2} -\lambda_{1}}  \ln cosh \lambda_{1}   +
       \ln cosh (\lambda_{2}+\zeta^{*})  \}
      \;\;\; ,
\end{eqnarray}
such that $\lambda_{2} -\lambda_{1}+\zeta^{*} \neq 0$.
{}From this relation I obtain
\begin{equation}\label{eq:Mat5}
D_{\zeta^{*}\rightarrow 0} = \frac{\tau (0)}{\lambda_{2} -\lambda_{1}}
           \;\;\; ,
\end{equation}
and
\begin{eqnarray}\label{eq:cc2}
D_{\zeta^{*}\rightarrow + \infty}
  & = & D_{\zeta^{*}\rightarrow 0} \pm
     \mu  \phi_{0}^{*} \{ 1 +\Gamma_{\lambda} \}  \;\;\; ,
         \nonumber \\
D_{\zeta^{*}\rightarrow -\infty}
  & = & D_{\zeta^{*}\rightarrow 0} \mp
     \mu  \phi_{0}^{*} \{ 1 -\Gamma_{\lambda} \}  \;\;\; .
\end{eqnarray}
However if $\lambda_{2} -\lambda_{1}+\zeta^{*}= 0$ then
\begin{eqnarray}\label{eq:Madd10}
D_{\zeta^{*}\rightarrow (\frac{-1}{\lambda_{2} -\lambda_{1} })}
  & = & D_{\zeta^{*}\rightarrow 0} \pm \mu \phi_{0}^{*} \{
    \Gamma_{\lambda}  + tanh \lambda_{1} \} \;\;\; .
\end{eqnarray}

Complementary to Eq.(\ref{eq:cca1}) I also define the following
dimensionless function
\begin{equation}\label{eq:cc1}
\alpha (\zeta^{*})  \equiv  \frac{\partial }{\partial \zeta^{*}}\tau
(\zeta^{*})
     =D_{\zeta^{*}\rightarrow 0} \pm \mu \phi_{0}^{*} \{
    \Gamma_{\lambda}  + tanh (\lambda_{2}+\zeta^{*}) \}   \;\;\; .
\end{equation}
Therefore, using Eq.(\ref{eq:cc1}) in conjunction with Eq.(\ref{eq:cc2}),
it can be easily shown that
\begin{eqnarray}\label{eq:bb1}
\alpha_{max}\equiv \alpha (\zeta^{*}\rightarrow-\infty)  & = &
    D_{\zeta^{*}\rightarrow -\infty}
      \;\;\; , \nonumber \\
\alpha_{min}\equiv \alpha (\zeta^{*}\rightarrow+\infty)   & = &
    D_{\zeta^{*}\rightarrow +\infty}  \;\;\; .
\end{eqnarray}
In order to relate these equations to the principle of MF
\cite{Lee88,Blum89}, I consider the following condition
for $\alpha$ and $D$.  If $\zeta^{*}\rightarrow + \infty$, then
$D_{\zeta^{*}\rightarrow +\infty}$ and $\alpha (\zeta^{*})$ are allow
to take one of the two values, namely 0 or $2D_{\zeta^{*}\rightarrow 0}$,
depending on the phenomena to be considered.  The particular case for which
$\alpha (\zeta^{*}\rightarrow +\infty )\approx 0$ and
$D_{\zeta^{*}\rightarrow +\infty}\approx 0$ will be in correspondence
with the features of MF.  To achieve this I approximate
\begin{equation}\label{eq:Mat3}
\mp\mu\phi_{0}^{*}\approx \frac{\varepsilon D_{\zeta^{*}\rightarrow 0}}
   { (1+ \Gamma_{\lambda}) }      \;\;\; ,
\end{equation}
where the integer factor $\varepsilon \equiv \pm 1$ is included to
distinguish these two limiting values.

According to the original definitions used in MF \cite{Hav89,Schw90}
I also define the dimensionless functions
\begin{equation}\label{eq:MMa1}
f(\alpha)  \equiv  \zeta^{*}\alpha (\zeta^{*})-\tau (\zeta^{*})   \;\;\; ,
\end{equation}
and
\begin{equation}\label{eq:Madd1}
C_{\zeta^{*}}  \equiv
    -\; \frac{\partial^{2}}{\partial (\zeta^{*})^{2}} \tau (\zeta^{*})
  = \mp \mu\phi_{0}^{*} \; sech^{2} (\lambda_{2}+\zeta^{*})  \;\;\; .
\end{equation}
When $f(\alpha )$ and $D_{\zeta^{*}}$ are smooth functions of $\alpha$ and
$\zeta^{*}$, then $f(\alpha )$ can be related to $\tau (\zeta^{*})$ by
a Legendre transformation \cite{Chh89}.  As is well known this reflects a
deep connection with the thermodynamic formalism of equilibrium statistical
mechanics where $\tau$ and $C_{\zeta^{*}}$ are the analogous `free energy'
and 'specific heat', respectively.  Within this
characterization $\alpha$ becomes an analogous `internal energy' and
$f$ an analogous `entropy' \cite{Hav89}.

Having established a physical meaning for these functions and
assuming $\phi_{0}^{*}>0$ as discussed below, I realize then that
$\mu$ can take the values
\begin{equation}\label{eq:Madd2}
\mu \rightarrow \mp \; 1  \;\;\; ,
\end{equation}
since it is physically satisfactory to set $C_{\zeta^{*}}\geq 0$.
This, in turn, implies that the reduced probability density
$\phi (\zeta )/\phi_{0}$ of Eq.(\ref{eq:can2a}) can take the desired
values $0$ and $1$ when $\zeta^{*}\rightarrow +\infty$ and
$\zeta^{*}\rightarrow -\infty$, respectively.  Within these
limits, in the case of MF,
the quantities $\alpha (\zeta^{*})$ of Eq.(\ref{eq:cc1}),
$D_{\zeta^{*}}$ of Eq.(\ref{eq:Mat1}) and $C_{\zeta^{*}}$ of
Eq.(\ref{eq:Madd1}) converge to zero whereas $\tau (\zeta^{*})$
of Eq.(\ref{eq:w1}) increases or decreases monotonically dependending
on $\zeta^{*}$.

Besides this, the choice of Eq.(\ref{eq:Madd2}) implies that
I may also obtain Eq.(\ref{eq:Mat3}) from Eq.(\ref{eq:Mat5}) in
conjuction with Eq.(\ref{eq:Mat2}) provided that $\varepsilon \rightarrow +1$.
However, in the case that $\varepsilon \rightarrow -1$ is choosen,
Eq.(\ref{eq:Mat3}) can only be recovered using again the
aformentioned equations but changing the sign of the previous value
for $\tau (0)$.  This will
became more clear latter in the calculations.  The present continuous
probability theory is thus dependent on $\lambda_{i}$ ($i$=1,2) and (the
sign and magnitude of) $\tau (0)$ in Eq.(\ref{eq:Mat2}),
where the reduced variable $\zeta$ satisfies the condition given by
Eq.(\ref{eq:Ant1}) and $\phi_{0}^{*}$ is positive satifying Eq.(\ref{eq:Mat3}).

{\em The case $\lambda_{1}>\lambda_{2}$ such that $\lambda_{2}>0$} (MF):
In Fig.1(a) I display the dependence of the analogous
`free energy' $\tau$ on the reduced coordinate variable $\zeta^{*}$ for
different values of $\lambda_{1}$ and $\lambda_{2}$, such that $\tau (0)=-1$
and $\varepsilon =1$.  Noting that the difference between
$\lambda_{2}$ and $\lambda_{1}$ is $-1$ in all three curves
illustrated, so as to have $D_{\zeta^{*}\rightarrow 0}=1$, then
from Eq.(\ref{eq:Ant1}) $\zeta \leq 1$.
Hitherto, the present GL-based approch to MF allows $\zeta^{*}$ to take on
negative as well as positive values $\leq 1$.  However, in the following plots
concerning MF, the range of $\zeta^{*}$ is extended to $3$ for illustrative
purposes.

{}From Fig.1(a) it can be seen that, on increasing the value of $\lambda_{1}$,
there is a more rapid convergence of $\tau (\zeta^{*})$ for positive
$\zeta^{*}$ than within the region $\zeta^{*}<0$; displaying thus typical
features of MF \cite{Hav89}.  Such a nontrivial behaviour
of $\tau$ illustrates the data collapse or breakdown of MF at
$\zeta^{*}>0$ where $\tau (\zeta^{*}) >0$.  This is in accordance with the
MF structure of the function $D_{\zeta^{*}}$ shown in Fig.2.
In fact, this corresponds to a spectrum of fractal dimensions.
In this figure I point out that $D_{\zeta^{*}}$ -defined via
Eq.(\ref{eq:Mat1})- is not constant for positive and negative values of
$\zeta^{*}$ so that it may correspond to a multifractal dimension.  Hence,
$\tau$ becomes also a nonlinear function of $\zeta^{*}$.  In fact this
approach yields, {\em e.g}, for the case
$\lambda_{1}=1, \lambda_{2}=0$ (full line in Fig.1(a)) the values
$D_{\zeta^{*}\rightarrow 1}=0.421$ and
$D_{\zeta^{*}\rightarrow -\infty}=3.53$ using Eqs.(\ref{eq:Madd10})
and (\ref{eq:cc2}), respectively.

On the other hand, in Fig.3(a) I display the analogous 'internal energy'
$\alpha$ as a function of $\zeta^{*}$ for the same set of values of
$\lambda_{1}$ and $\lambda_{2}$ as in Fig.1(a).
When $\lambda_{1}=1$ and $\lambda_{2}=0$ this function exhibits
sharp variations around $\zeta^{*}=0$  with a maximum value that shows
a stronger dependence on $\lambda_{1}$ for positive $\zeta^{*}$ than for
negative $\zeta^{*}$.  This rules out the possibility that the actual
position for a critical value of $\zeta$ (or analogous critical inverse
'temperature' \cite{Hav89}),
at which the multifractal formalism actually breaks down, is obtained
at $\zeta_{c}^{*}=0$.  For other values of $\lambda_{1}$ and $\lambda_{2}$,
$\zeta^{*}_{c}$ changes towards negative values.
One should also bear in mind that for
$\zeta^{*}<\zeta^{*}_{c}$, the analogous 'free energy' $\tau (\zeta^{*})$ is
dominated by $\alpha$ which, in turn, varies with the magnitude of
$\lambda_{2}$.

Characteristic features of a phase transition at $\zeta^{*}_{c}$ can be
figure out by examining the shape of the analogous 'specific heat'
$C_{\zeta^{*}}$ of Eq.(\ref{eq:Madd1}) which is illustrated in Fig.4.
It can be easily visualized that there is a sharp peak around the value
$\zeta^{*}_{c}=0$ for the case corresponding to the full line in Fig.1(a).
The heights and positions of these curves are strongly dependent on
$\lambda_{1}>\lambda_{2}$.  Nicely, this finding is also similar to reported
MF results \cite{Hav89}.  Finally, I investigate the behaviour of the
analogous 'entropy' $f$ against $\alpha$ for several values of
$\lambda_{1}$ and $\lambda_{2}$.  As can be seen from Fig.5(a) I find that,
on incresing the magnitude of $\lambda_{1}$, the left-hand side
of these plots converge more rapidly than the right-hand sides which converge
poorly.  This is in complete agreement with the MF signal observed
in the context of self-similar random resistor networks
(open and full circles in Fig.5(a) from \cite{Naga89}) or, {\em e.g.}, in
DLA when $\lambda_{1}$ is taken to be related to the system size
of the simulation box \cite{Lee88,Schw90}.  In fact, this singularity
$f(\alpha )$ shows the characteristic normalized convex shape found in MF.
Given its interpretation of a multifractal character for fractal subsets,
each with a different fractal dimension having singularity strength
$\alpha$, it is expected that $f(\alpha )\geq 0$.

Let me remark again that all of these predictions resemble qualitatively
many of the intriguing results observed in MF, such as the breakdown
of multifractal behaviour and the existence of a phase transition
at $\zeta^{*}$, which may not necessarily be at $\zeta^{*}=0$.
In the examples considered above the maximum and minimum values of the
analogous 'internal energy' $\alpha$ can be estimated from Eqs.(\ref{eq:bb1})
in conjuction with Eq.(\ref{eq:cc2}), respectively.  This allows for the
existence of a critical point $\zeta^{*}$ above which the infinite hierarchy
(or broad distribution) of phases can be found, but below which a
single phase appears characterized by the maximum 'energy' $\alpha_{max}$.

{\em The case $\lambda_{1}>\lambda_{2}$ such that $\lambda_{2}<0$} (SOC):
In Fig.1(b) the dependence of the normalized probability
distribution function $\tau$ of Eq.(\ref{eq:w1}) on the $energy/unit force$
variable $\zeta^{*}$ is plotted for different values of
$\lambda_{1}$ and $\lambda_{2}$ such that, as a main difference with
respect to MF, $\tau (0)=1$ and $\varepsilon =-1$.
This new choice of $\tau(0)$ allows to normalize $\tau (\zeta^{*})$
and the choice of $\varepsilon$ enables to mimic the main features of SOC,
namely a power-law behaviour, provided the reduced variable $\zeta^{*}$ is
associated with the $log$-function of some measured event.  As an illustrative
example, in Fig.1(c) I show a reasonable description of the probability $\pi$
that the water flow intensity of a randomly chosen site is larger than $s$,
as obtained from the SOC signal calculated within a
model of erosion \cite{Taka92}.
This demostrates that the theory do apply to self-organizing systems.
For a class of continuous, cellular automaton models of earthquakes
\cite{Ola92,Car89}, the function $\tau$ is simply reinterpreted as being the
number of events with reduced released energy $E\sim e^{\zeta^{*}}$.

The several theoretical curves in Fig.1(b) refer to different values of
($\hspace{1.2cm}$)
$\lambda_{1}=3$ and $\lambda_{2}=-6$; ($- - - -$) $\lambda_{1}=3$,
$\lambda_{2}=-7$;
($\cdots\cdots$) $\lambda_{1}=3$, $\lambda_{2}=-8$.
Using Eq.(\ref{eq:Ant1}), then the GL-based theory is valid
for $\zeta \leq 9,10,11$, respectively, where I deal with the physically
interesting interval $0<\tau (\zeta^{*})\leq 1$.
A given slope of the linear behaviour of
the curves in this figure is determined by fixing $\lambda_{1}>0$.
In particular, this parameter may be associated with the elastic
parameter of the spring-block model for earthquakes which links the
rate of occurrence of
earthquakes of magnitude $M$ greater than $m$ to the energy (seismic moment)
$E$ released during the earthquake via the famous empirical Gutenber-Richter
law \cite{Ola92}.  And the cutoff in the axes $\zeta^{*}$ may be
related to the system size of cellular automaton modeling for
arrays of threshold elements.

To see more clearly power-law features in the behaviour of $\tau$ over
a wide range of positive values of $\zeta^{*}$ (as those in Figs.1(b) and
(c)), it is necessary to
investigate the behavior of the derivative of $\tau (\zeta^{*})$, defined
through $\alpha$ of Eq.(\ref{eq:cc1}), which is plotted in Fig.3(b).
A glance into the behaviour of $\alpha (\zeta^{*})$ indicates that
for the smallest displayed values of $\zeta^{*}>0$ the $\alpha$-function
converges to a constant negative value, revealing in this way the constant
nature of the negative slopes in the ($\tau$-$\zeta^{*}$) curves of
Fig.1(b).  On increasing $\zeta^{*}$ each curve smoothly approach
a smaller value.  Clearly, due to the probabilistic nature of $\tau$ for SOC
$(i)$ such convergences of $\alpha$ towards smaller negative values need not
to be considered and $(ii)$ the relations between $\alpha_{max,min}$ and
$D_{\zeta^{*}\rightarrow \mp\infty}$ ({\em c.f.}, Eq.(\ref{eq:bb1})),
become meaningless.

After establishing this resemblance of a power-law
description, it is tempting to continue applying anew
the above formalism of MF to analyse SOC.  As already mentioned, this
may help to understand the origin of SOC from the
novel viewpoint of a nonlinear singularity spectrum different from
what is common to multifractal objects ({\em c.f.}, Fig.5(a)).

In view of the features in $\alpha (\zeta^{*})$ of Fig.3(b), the second
derivative of $\tau$ (not shown), namely $C_{\zeta^{*}}$ of
Eq.(\ref{eq:Madd1}), present a sharp peak around
an inflection point of the function $\alpha (\zeta^{*})$ of Fig.3(b),
say $\zeta^{*}_{inf}$.  As a difference to the above results of MF
({\em c.f.}, Fig.4), these peak heights reduce their magnitude on decresing
$\lambda_{2}<\lambda_{1}$ and shift their position towards positive
values of $\zeta^{*}$.  In this case no phase transition as in the case of
MF is expected because $\tau >0$ restricts the range of valid
$\zeta^{*}<\zeta^{*}_{inf}$.  Moreover, as mentioned above,
$D_{\zeta^{*}}$ remains undefined for SOC due to the re-interpretation of
$\tau (\zeta^{*})$.  Similarly to what is shown in Fig.2 this function
for SOC may simply become not constant on increasing $\zeta^{*}$ as in MF.

On the other hand, the continuous singularity spectrum $f(\alpha )$
plays an alternative role when dealing with SOC which can be assessed from
Fig.5(b).  In this plot it can be seen that this function
exhibits a rather nonlinear behaviour different from the parabolic behaviour
typical of multifractal entities ({\em c.f.}, Fig.5(a)).  In MF
$f$ takes its maximum at the value $\alpha (\zeta^{*}=0)$
whereas in SOC this spectrum becames a monotonically increasing
(negative) function of (negative) $\alpha$.  On decresing the magnitude of
$\lambda_{2}$ these curves converge to $-1$ and separate out as a function
of decreasing $\alpha$ whithin the range of validity of $\alpha$ in Fig.3(b).
Since $\tau (\zeta^{*})$ is positive, then $-2<f(\alpha )<-1$.
I suggest this new aspect of the behaviour of
$f(\alpha )$ to be a fundamental property for the additional characterization
of SOC.  It is therefore most likely that the linear behaviour displayed by
$\tau (\zeta^{*})$ ({\em c.f.}, Fig.1(b)), that is quantified via
$\alpha (\zeta^{*})$ ({\em c.f.}, Fig.3(b)), finds its
root through the behaviour of Eq.(\ref{eq:MMa1}) for $f(\alpha )$ ({\em c.f.},
Fig. 5(b)).

To some extent, I have been able to shed light on a basic mechanism
leading to both concepts of MF and SOC from the continuous density probability
$\phi (\zeta )$ as given in Eq.(\ref{eq:can2a}).  This function
has been assumed to be related to $H(\zeta )$ via Eq.(\ref{eq:can2a}),
which I postulated to be given by the real kink
solutions of a dimensionless GL-like equation.  Such solutions
are known to minimize the GL free energy functional
when $\mid \phi \mid^{2}$ is related to a particle concentration \cite{Gin50}.
Of course to use a continuous probability theory may be seen as being
heuristic, specially so if reported simulation results have been done using
discretized cell configurations.  But, as I have discussed, a great
deal of relevant information can be extracted from a continuos approach which,
essentially, does relay on the sign of $\lambda_{2}$, $\tau (0)$ and
$\varepsilon$ only.

The thermodynamic treatment of MF provides a convenient
way to quantify the complexity of multifractals by characterizing
scale-invariant singularities and, in most cases, allowing
for an effective comparison between theory and experiments
\cite{Lee88,Blum89,Hav89,Schw90}.  Nevertheless, I have proposed here
that this formalism may be extended as a tool to rationalize SOC
on the basis of a singularity spectrum that turns out to be nonlinear
as well (when comparing Figs.5(a) and (b)).
While the present (static) GL-based theory is extremely simple, it is
important to emphasize that it gives information about the complex origin
of self-organized critical phenomena whose physics is assumed -to a good
approximation- to be analog to that required to describe MF.  This theory
also reflects the minimal ingredients than can give rise
to an intrinsically critical state. In passing,
I add that multifractal structures can also emerge from
GL equations with random initial conditions for the
temporal evolution \cite{Brax92}.

It is a pleasure to acknowledge helpful remarks by Drs. V.L. Nguyen,
G.C. Barker, W. Wang and H. Rosu.

\newpage

\section*{Figure captions}

\begin{itemize}

\item {\bf Figure 1}: (a) Dependence of the analogous `free energy'
$\tau$ on the coordinate variable $\zeta^{*}$ displaying typical
features of MF phenomena \cite{Lee88}.  Different values of ($\hspace{1.2cm}$)
$\lambda_{1}=1$ and $\lambda_{2}=0$; ($- - - -$) $\lambda_{1}=1.5$,
$\lambda_{2}$=0.5; ($\cdots\cdots$) $\lambda_{1}=2$, $\lambda_{2}=1$,
illustrates the data collapse at $\zeta^{*}>0$ where $\tau >0$.
In these examples $\tau (0)=-1$ and $\varepsilon =1$.

(b) Dependence of the normalized probability distribution function
$\tau$ on the variable $\zeta^{*}$ displaying typical features of SOC.
Different curves refer
to different values of ($\hspace{1.2cm}$) $\lambda_{1}=3$ and $\lambda_{2}=-6$;
($- - - -$) $\lambda_{1}=3$, $\lambda_{2}=-7$; and ($\cdots\cdots$)
$\lambda_{1}=3$, $\lambda_{2}=-8$.  In these examples $\tau (0)=1$ and
$\varepsilon =-1$.

(c) Present description of the probability $\pi$
that the water flow intensity of a randomly chosen site is larger than
the event $s$.  Open circles are the SOC signal in a model of
erosion \cite{Taka92}.

\vspace{1cm}

\item {\bf Figure 2}: Plot of the multifractal dimension
$D_{\zeta^{*}}$ against $\zeta^{*}$ showing a non-constant behaviour
for the same set of values for $\lambda_{1}$ and
$\lambda_{2}$ as in Fig.1(a).

\vspace{1cm}

\item {\bf Figure 3}: (a) Dependence of the analogous 'internal energy'
$\alpha$ on $\zeta^{*}$ displaying characteristic features of a phase
transition for the same set of values for $\lambda_{1}$ and $\lambda_{2}$
as in Fig.1(a).

(b) The function $\alpha$ of Eq.(\ref{eq:cc1}) as a function of $\zeta^{*}$
for the same set of values for $\lambda_{1}$ and $\lambda_{2}$
as in Fig.1(b).  These curves reflect the power-law features in the behaviour
of $\tau (\zeta^{*})$ for SOC.

\vspace{1cm}

\item {\bf Figure 4}:  Analogous 'specific heat' $C_{\zeta^{*}}$ as a
function of $\zeta^{*}$ displaying characteristic features of a phase
transition for the same set of values for $\lambda_{1}$ and $\lambda_{2}$
as in Fig.1(a).

\vspace{1cm}

\item {\bf Figure 5}: (a) Dependence of the analogous 'entropy' $f$ on the
analogous 'internal energy' $\alpha$ for the same set of values for
$\lambda_{1}$
and $\lambda_{2}$ as in Fig.1(a).  Open and full circles are the MF signal
calculated in the context of self-similar random resistor networks
for different cell sizes (see \cite{Naga89}).

(b) The nonlinear $f(\alpha )$ singularity spectrum obtained from
Eq.(\ref{eq:MMa1}) proposed to characterize SOC using the same set
of values for $\lambda_{1}$ and $\lambda_{2}$ as in Fig.1(b).

\end{itemize}


\newpage

\begin{thebibliography}{99}

\bibitem{Bak87} P. Bak, C. Tang and K. Weisenfeld, {\em Phys. Rev. Lett}
{\bf 59} (1987) 381.
\bibitem{Lee91} S.-C. Lee, N.Y. Liang and W.-J. Tzeng, {\em Phys. Rev. Lett.}
{\bf 67} (1991) 1479.
\bibitem{Gri90} G. Grinstein, D.-H. Lee and S. Sachdev, {\em Phys. Rev. Lett}
{\bf 64} (1990) 1927.
\bibitem{Dio91} P. Diodati, F. Marchesoni and S. Piazza, {\em Phys. Rev. Lett}
{\bf 67} (1991) 2239.
\bibitem{Cot91} P.J. Cote and L.V. Meisel, {\em Phys. Rev. Lett.} {\bf 67}
(1991) 1334.
\bibitem{Am86} C. Amitrano, A. Coniglio and F. di Liberto,
{\em Phys. Rev. Lett.} {\bf 57} (1986) 1016.
\bibitem{Lee88} J. Lee and H. E. Stanley, {\em Phys. Rev. Lett.} {\bf 61}
(1988) 2945; {\em Ibid} {\bf 63} (1989) 1190.
\bibitem{Mut83} M. Muthukamar, {\em Phys. Rev. Lett.} {\bf 50} (1983) 839.
\bibitem{Gou83} H. Gould, F. Family and H.E. Stanley, {\em Phys. Rev. Lett.}
{\bf 50} (1983) 696.
\bibitem{Chh89} A. Chhabra and R.V. Jensen, {\em Phys. Rev. Lett.}
{\bf 62} (1989) 1327.
\bibitem{Piwr91} L. Pietronero and W.R. Schneider, {\em Phys. Rev. Lett.}
{\bf 66} (1991) 2336.
\bibitem{Gin50} V.L. Ginzburg and L.D. Landau, {\em Zh. Eksp. Teor. Fiz.}
{\bf 20} (1950) 1064.
\bibitem{Fel66} W. Feller, in {\em `An introduction to probability theory
and its applications'}, Vol.2 (Wiley, N.Y. 1966).
\bibitem{Blum89} R. Blumenfeld and A. Aharony, {\em Phys. Rev. Lett.}
{\bf 62} (1989) 2977.
\bibitem{Hav89} S. Havlin, B. Trus, A. Bunde and H. E. Roman,
{\em Phys. Rev. Lett.} {\bf 63} (1989) 1189.
\bibitem{Schw90} S. Schwarzer, J. Lee, A. Bunde, S. Havlin, H.E. Roman
and H.E. Stanley, {\em Phys. Rev. Lett.} {\bf 65} (1990) 603.
\bibitem{Naga89} T. Nagatani, M. Ohki and M. Hori, {\em J. Phys. A: Math. Gen.}
{\bf 22} (1989) 1111.
\bibitem{Taka92} H. Takayasu and H. Inaoka, {\em Phys. Rev. Lett.} {\bf 68}
(1992) 966.
\bibitem{Ola92} Z. Olami, H.J.S. Feder and K. Christensen,
{\em Phys. Rev. Lett.} {\bf 68} (1992) 1244.
\bibitem{Car89} J.M. Carlson and J.S. Langer, {\em Phys. Rev. Lett.}
{\bf 62} (1989) 2632.
\bibitem{Brax92} P. Brax, {\em Phys. Lett. A} {\bf 165} (1992) 335.

\end{thebibliography}
\end{document}